\documentclass[10pt,fleqn]{article}
\usepackage{amsmath}
\usepackage{amssymb}
\usepackage{amsfonts}
\usepackage{espcrc1}
\usepackage[latin1]{inputenc}
\usepackage[top=2.5cm, left=2.5cm, right=2cm, bottom=2.5cm]{geometry}

\input{tcilatex}
\begin{document}

\title{\LARGE The canonical structure of Podolsky's generalized electrodynamics on the Null-Plane}%

\author{M. C. Bertin\address[IFT]{Instituto de F\'{i}sica Te\'{o}rica - S\~{a}o Paulo State University.
Rua Dr. Bento Teobaldo Ferraz, 271 - Bloco II - Barra Funda, 01140-070, S\~ao Paulo, SP, Brazil}\thanks{mcbertin@ift.unesp.br.},\
B. M. Pimentel\addressmark[IFT]\thanks{pimentel@ift.unesp.br.},\
G. E. R. Zambrano\addressmark[IFT]\thanks{gramos@ift.unesp.br.}\thanks{On leave of absence from Departamento de F\'{i}sica, Universidad de Nari\~{n}o, San Juan de Pasto, Nari\~{n}o, Colombia.}}%

\maketitle%
\thispagestyle{empty}

\begin{abstract}%

In this work we will develop the canonical structure of Podolsky's
generalized electrodynamics on the null-plane. This theory has second-order
derivatives in the Lagrangian function and requires a closer study for the
definition of the momenta and canonical Hamiltonian of the system. On the
null-plane the field equations also demand a different analysis of the
initial-boundary value problem and proper conditions must be chosen on the
null-planes. We will show that the constraint structure, based on Dirac
formalism, presents a set of second-class constraints, which are exclusive
of the analysis on the null-plane, and an expected set of first-class
constraints that are generators of a $U\left( 1\right) $ group of gauge
transformations. An inspection on the field equations will lead us to the
generalized radiation gauge on the null-plane, and Dirac Brackets will be
introduced considering the problem of uniqueness of these brackets under the
chosen initial-boundary condition of the theory.

\end{abstract}%

\section{Introduction}

Most physical systems, including fundamental fields in quantum field theory,
are described by Lagrangians that depend at most on first-order derivatives.
However, there is a continuous interest on theories with higher-order
derivatives, either do accomplish generalizations or to get rid of some
undesirable properties of first-order theories. This interest had begun in
the half of the 19th century, when Ostrogradski \cite{ostro} developed the
Hamiltonian formalism for this kind of system in classical mechanics.

As examples of systems treated by higher-order Lagrangians we mention the
attempts to solve the problem of renormalization of the gravitational field
by inserting quadratic terms of the Riemann tensor and its contractions \cite%
{st0,st,qu} on the Einstein-Hilbert Action. Recent developments in this
direction has been made by Cuzinatto \emph{et al.} \cite{1a} where the
construction of high-order Lagrangians for gravity is made with invariants
of the Riemann tensor taking account the local Lorentz invariance. This
attempt turns out to be a natural generalization of the Utiyama's Theory of
General Gauge Fields \cite{uti} applied to second-order theories \cite{1b}.

Higher-order Lagrangians have also emerged as effective theories on the
infrared sector of the QCD \cite{qcd}, where it enforces a good asymptotic
behavior of the gluon propagator. It is also important to remark that the
inclusion of higher-order derivatives in field theory of supersymmetric
fields has shown to be a powerful regularization mechanism \cite{s1,s2}. We
noted that a very attractive property of quantum field theories with
higher-order terms is the fact that it improves the convergence of the
corresponding Feynman diagrams\cite{nes,al}.

The first model of a higher-order derivative field theory is a
generalization of the electromagnetic field proposed in the works of
Podolsky, Schwed and Bopp \cite{4,5}, which culminated in the Podolsky's
generalized electrodynamics. It is suggested to modify the Maxwell-Lorentz
theory in order to avoid divergences such as the electron self-energy and
the vacuum polarization current. These difficulties can be traced to the
fact that the classical electrodynamics involve an $r^{-1}$ singularity
that\ results\ in an infinite value of the electron self-energy. The
Lagrangian density is, therefore, modified by a second-order derivative term:%
\begin{equation}
\mathcal{L}=-\frac{1}{4}F_{\mu \nu }F^{\mu \nu }+\frac{1}{2}a^{2}\partial _{\lambda
}F^{\mu \lambda }\partial ^{\gamma }F_{\mu \gamma }\ ,\hspace{2cm}F_{\mu \nu
}\equiv \partial _{\mu }A_{\nu }-\partial _{\nu }A_{\mu }\ .  \label{I1}
\end{equation}%
Podolsky's theory already has many interesting features at the classical
level. It solves the problem of infinite energy in the electrostatic case
and also gives the correct expression for the self-force of charged
particles at short distances, as showed by Frenkel \cite{6}, solving the
problem of the singularity at $\mathbf{r}\rightarrow 0$. It has been shown
by Cuzinatto \emph{et al.} \cite{7} that the above Lagrangian density is the
only possible generalization of the electromagnetic field that preserves
invariance under $U(1)$. Besides, the theory yields field equations that are
still linear in the fields.

Another important prediction of the model is the existence of massive
photons, whose mass is proportional to the inverse of the Podolsky's
parameter $a$. This feature allows experiments that may test the generalized
electrodynamics as a viable effective theory. The determination of an upper
bound value for the mass of the photon is actually a current concern in the
theoretical framework \cite{7}.

The canonical quantization of the field was tried in the work of Podolsky
and Schwed \cite{4}. However, Podolsky's theory inherits the same
difficulties from the standard electromagnetic field, the presence of a
degenerate variable, which forced them to use a Fermi-like Lagrangian.
However, the chosen gauge fixing condition, the usual Lorenz condition, does
not fulfill the requirements for a good choice of gauge in the context of
Podolsky's theory. The first consistent approach to the quantization of the
field was given by Galv\~{a}o and Pimentel \cite{8}, where Dirac canonical
formalism \cite{dirac,dirac1,dirac2}\ is used with the correct choice of
gauge.

The first attempts of quantization of Podolsky's field was made in
instant-form, where the \textquotedblleft laboratory time\textquotedblright\
$t=x^{0}$ is the evolution parameter of the theory. Dirac \cite{dirac3} was
the first to notice that this choice is not the only possible
parametrization for field theories. Actually it is possible to define five
different forms of Hamiltonian dynamics, each one related to different
sub-groups of the Poincar\'{e} group \cite{bekker}. In this work we intend to
proceed with the canonical analysis in front-form dynamics, also called the
null-plane parametrization, where the coordinate $x^{+}\equiv 1/\sqrt{2}%
\left( x^{0}+x^{3}\right) $ is chosen to be the evolution parameter. This
parameter choice was mistaken, for some time, to the so called
infinite-momentum frame \cite{frame}, which is a limit process to analyse
field theories in a frame near to the speed of light. It is a choice of
coordinate system rather than a physical reference frame, and implies the
definition of the null-plane which is the 3-surface $x^{+}=0$. The classical
evolution of the system is then given by the definition of appropriate
brackets plus a set of initial data, which are the configuration of the
fields at the above 3-surface.

The paper is organized as follows. In section two we discuss the null-plane
coordinates, which are the natural coordinate system for the front-form
dynamics, and we review the initial-boundary value problem for the fields
and establish appropriate conditions to achieve a unique solution of the
dynamic equations on the null-plane. Section three will be devoted to a
review on the Hamiltonian formalism for higher-order Lagrangians. In Section
four the canonical approach is applied for the generalized electromagnetic
field in null-plane coordinates. In section five we establish a set of
consistent gauge conditions and corresponding Dirac Brackets to describe the
physical dynamics of the theory. Section six will be devoted for the final
remarks.

\section{The null-plane coordinates}

As the start-point for the analysis of a field theory we have the Action%
\begin{equation*}
S\left[ \phi \right] \equiv \int_{\Omega }d\omega \ \mathcal{L}\left( \phi
,\partial _{\mu }\phi \right) \ ,
\end{equation*}%
where $\mathcal{L}$ is a Lagrangian density, and $d\omega $ is a four-volume
element of a finite (or usually infinite) four-volume $\Omega $ of the
space-time. For relativistic theories the Lagrangian density must be chosen
to be invariant under any particular parameter choice. However, although the
Lagrangian formalism preserves this invariance, the same does not occur in
the Hamiltonian formalism, which requires a parametrization in order to be
fully carried out.

Dirac has shown \cite{dirac3} that the usual dynamics, the instant-form, where the galilean time $x^{0}=t$ is the parameter that defines the
evolution of the system from a given initial 3-surface $\Gamma _{t=t_{0}}$
to a later surface $\Gamma _{t=t_{1}}$, is not the only possible choice of
parametrization. He calls attention for two other forms of Hamiltonian
dynamics: the punctual-form and the front-form. Later, two other forms were
discovered \cite{forms}.

An important advantage pointed out by Dirac is the fact that seven of the
ten Poincar\'{e} generators are kinematical on the null-plane while the
conventional theory constructed in instant-form has only six of these
generators. Therefore, the structure of the phase space is distinct in both
cases. As such, a description of the physical systems on the null-plane
could give additional information from those provided by the conventional
formalism \cite{sri}. Another remarkable feature is that regular theories
become constrained when analyzed on the null-plane. In general, it leads to
a reduction in the number of independent field operators in the respective
phase space due to the presence of second-class constraints.

The natural coordinate system of instant-form dynamics is the rectangular
system $x^{\mu }\equiv \left( x^{0},x^{1},x^{2},x^{3}\right) $. We can pass
to null-plane coordinates with the linear transformation $x^{\prime }=\Sigma
\ x$ where the transformation matrix and its inverse are given by%
\begin{equation*}
\Sigma \equiv \frac{1}{\sqrt{2}}\left(
\begin{array}{ccc}
1 & 0 & 1 \\
1 & 0 & -1 \\
0 & \sqrt{2}\cdot \mathbf{I} & 0%
\end{array}%
\right) \ ,\hspace{1cm}\Sigma ^{-1}=\frac{1}{\sqrt{2}}\left(
\begin{array}{ccc}
1 & 1 & 0 \\
0 & 0 & \sqrt{2}\cdot \mathbf{I} \\
1 & -1 & 0%
\end{array}%
\right) \ ,
\end{equation*}%
where $\mathbf{I}$ is the $2\times 2$ identity matrix and $x^{\prime \mu
}\equiv \left( x^{+},x^{-},x^{1},x^{2}\right) $.

Lorentz tensors are also covariant under this transformation, but the
transformation itself is not of Lorentz type \cite{sri}: if in usual
coordinates we define the Minkowski metric as $\eta \equiv $diag$\left(
1,-1,-1,-1\right) $, the metric in null-plane coordinates will be given by%
\begin{equation*}
\eta ^{\prime }=\Sigma \ \eta \ \Sigma ^{-1}=\left(
\begin{array}{ccc}
0 & 1 & 0 \\
1 & 0 & 0 \\
0 & 0 & -\mathbf{I}%
\end{array}%
\right) \ .
\end{equation*}%
Using this metric (from now on we will ignore the comma) we can see that the
norm of a vector is not a quadratic form, but will be linear in the
longitudinal\ components.

Of special interest is the D'Alambertian operator%
\begin{equation}
\square \equiv \partial _{\mu }\partial ^{\mu }=2\partial _{-}\partial
_{+}+\partial _{i}\partial ^{i}\ .  \label{dalambertian}
\end{equation}%
Since the evolution parameter is $x^{+}$ a field equation like $\left(
\square +m^{2}\right) \phi =0$ will be linear on the velocity $\partial
_{+}\phi $, which does not occur in instant-form. Therefore, the analysis of
initial-boundary value problem is changed from a Cauchy to a characteristics
initial-boundary value problem. This is due to the fact that a quadratic Lagrangian on $\partial _{0}\phi $ is actually of first-order on $\partial _{+}\phi $. In the case of the scalar field on
the null-plane it is sufficient to fix the values of the fields on both
characteristics surfaces to solve the field equations \cite{nev,roh,wer}.

This can be seen in Podolsky's case by the Euler-Lagrange (EL) equations of
the Lagrangian (\ref{I1}):%
\begin{equation}
\left( 1+a^{2}\square \right) \square A_{\mu }-\partial _{\mu }\left(
1+a^{2}\square \right) \partial ^{\nu }A_{\nu }=0\ .  \label{ELPodolsky}
\end{equation}%
This equation is fourth-order in $\partial _{0}A_{\mu }$ but only
second-order in $\partial _{+}A_{\mu }$. Therefore, in instant-form it is
necessary to specify four conditions, the values of the field and of its
derivatives until third-order on an initial surface $x^{0}=0$ to uniquely
write a solution.

On the null-plane the equation is just of second-order, but the existence of
two characteristics surfaces demands the knowledge of four initial-boundary
conditions as well. The normal vector of a null-plane lies in the same
plane, therefore, the knowledge of the value $A_{\mu }$ on a null-plane
implies in its normal derivative $\partial _{+}A_{\mu }$. Thus, the solution
of the field equations is uniquely determined if $A$ is specified on the
null-plane $x^{+}=cte$ and three boundary conditions are imposed on $%
x^{-}=cte$, which in our case consists on the value of the derivatives of
the field up to third-order.

In the canonical framework, it was Steinhardt \cite{ste} who showed that to
linear Lagrangians the initial condition on $x^{+}=cte$ plus a Hamiltonian
function are insufficient to predict uniquely all physical process. Boundary
conditions along the $x^{-}=cte$ plane must also be determined. He also
observed that the matrix formed by the Poisson Brackets (PB) of the
second-class constraints does not have a unique inverse and that the
presence of arbitrary functions is associated with the insufficiency of the
initial value data. It is also responsible for the existence of a hidden
subset of first-class constraints which is associated with improper gauge
transformations \cite{BCT}. By imposing appropriate initial-boundary
conditions on the fields, the hidden first-class constraints can be
eliminated in order to the total Hamiltonian be a true generator of the
physical evolution. It will also determine an unique inverse of the second
class constraint matrix which allows to obtain the correct Dirac Brackets
among the fundamental variables. Thus, in the study of the Podolsky's theory
we follow the same tune outlined in \cite{Cas,9,10}.

\section{Second-order derivatives \textbf{on the null-plane}}

Let us consider a generic Lagrangian density $\mathcal{L}(\phi ,\partial
\phi ,\partial ^{2}\phi )$ dependent of a number $n$ of fields $\phi ^{a}(x)$
and its first and second derivatives. The application of Hamilton's Action
principle yields the following EL equations%
\begin{equation*}
\frac{\delta \mathcal{L}}{\delta \phi ^{a}}-\partial _{\mu }\left[ \frac{%
\delta \mathcal{L}}{\delta \left( \partial _{\mu }\phi ^{a}\right) }\right]
+\partial _{\mu }\partial _{\nu }\left[ \frac{\delta \mathcal{L}}{\delta
\left( \partial _{\mu }\partial _{\nu }\phi ^{a}\right) }\right] =0\ ,
\end{equation*}%
which are the equations that originates (\ref{ELPodolsky}) from (\ref{I1}).
On the solutions, the vanishing of the variation of the action yields the
conserved symmetric energy-momentum tensor%
\begin{eqnarray}
T_{\mu \nu } &\equiv &\partial _{\mu }\phi ^{a}\ \frac{\delta \mathcal{L}}{%
\delta \left( \partial ^{\nu }\phi ^{a}\right) }-\mathcal{L}\eta _{\mu \nu }
\notag \\
&&-2\partial _{\mu }\phi ^{a}\partial _{\lambda }\left[ \frac{\delta
\mathcal{L}}{\delta \left( \partial ^{\nu }\partial _{\lambda }\phi
^{a}\right) }\right] +\partial _{\lambda }\left[ \partial _{\mu }\phi ^{a}\
\frac{\delta \mathcal{L}}{\delta \left( \partial ^{\nu }\partial _{\lambda
}\phi ^{a}\right) }\right] -\partial ^{\lambda }\left( \Xi _{\mu \lambda \nu
}+\Pi _{\mu \lambda \nu }\right) \ ,  \label{EMT}
\end{eqnarray}%
where%
\begin{eqnarray*}
\Xi _{\mu \lambda \nu } &\equiv &\frac{1}{2}\left[ \frac{\delta \mathcal{L}}{%
\delta \left( \partial ^{\mu }\phi ^{a}\right) }-\partial _{v}\left( \frac{%
\delta \mathcal{L}}{\delta \left( \partial ^{\mu }\partial _{\nu }\phi
^{a}\right) }\right) \right] \left( \mathbf{I}_{\lambda \nu }\right) _{\
b}^{a}\ \phi ^{b}\ , \\
\Pi _{\mu \lambda \nu } &\equiv &\frac{1}{2}\frac{\delta \mathcal{L}}{\delta
\left( \partial ^{\mu }\partial ^{\alpha }\phi ^{a}\right) }\ \left( \mathbf{%
I}_{\lambda \nu }\right) _{\ b}^{a}\ \partial ^{\alpha }\phi ^{b}\ .
\end{eqnarray*}%
$\left( \mathbf{I}_{\lambda \nu }\right) _{\ b}^{a}$ are the infinitesimal
generators of the Poincar\'{e} group.

The conserved charge is given by the expression%
\begin{equation*}
G\equiv -a^{\mu }P_{\mu }-\frac{1}{2}\omega ^{\mu \nu }M_{\mu \nu }
\end{equation*}%
with generators%
\begin{eqnarray*}
P_{\mu } &\equiv &\int_{\sigma }d\sigma ^{\nu }T_{\mu \nu }\ , \\
M_{\mu \nu } &\equiv &\int_{\sigma }d\sigma ^{\alpha }\left( T_{\alpha \mu
}x_{\nu }-T_{\alpha \nu }x_{\mu }\right) \ .
\end{eqnarray*}%
In the above expressions $\sigma $ is a 3-surface orthogonal to the
parametrization axis.

If we choose the null-plane, we will be interested in the dynamical
generator $P_{+}$, which is given by%
\begin{equation*}
P_{+}\equiv \int d^{3}xT_{+-}
\end{equation*}%
where we adopt $d^{3}x\equiv dx^{-}dx^{1}dx^{2}$. Here we have the canonical
Hamiltonian density%
\begin{eqnarray}
\mathcal{H}_{c} &\equiv &T_{+-}=\left[ \frac{\delta \mathcal{L}}{\delta
\left( \partial _{+}\phi ^{a}\right) }-\partial _{+}\ \frac{\delta \mathcal{L%
}}{\delta \left( \partial _{+}\partial _{+}\phi ^{a}\right) }-2\partial
_{-}\ \frac{\delta \mathcal{L}}{\delta \left( \partial _{-}\partial _{+}\phi
^{a}\right) }-2\partial _{i}\ \frac{\delta \mathcal{L}}{\delta \left(
\partial _{i}\partial _{+}\phi ^{a}\right) }\right] \partial _{+}\phi ^{a}
\notag \\
&&+\partial _{+}\partial _{+}\phi ^{a}\ \frac{\delta \mathcal{L}}{\delta
\left( \partial _{+}\partial _{+}\phi ^{a}\right) }-\mathcal{L}\ .
\label{energy}
\end{eqnarray}%
This result suggests the following definition for the canonical momenta:
\begin{subequations}
\label{momenta}
\begin{eqnarray}
p_{a} &\equiv &\int d^{3}x\left[ \frac{\delta \mathcal{L}}{\delta \left(
\partial _{+}\phi ^{a}\right) }-\partial _{+}\ \frac{\delta \mathcal{L}}{%
\delta \left( \partial _{+}\partial _{+}\phi ^{a}\right) }-2\partial _{-}\
\frac{\delta \mathcal{L}}{\delta \left( \partial _{-}\partial _{+}\phi
^{a}\right) }-2\partial _{i}\ \frac{\delta \mathcal{L}}{\delta \left(
\partial _{i}\partial _{+}\phi ^{a}\right) }\right] \ ,  \label{p} \\
\pi _{a} &\equiv &\int d^{3}x\frac{\delta \mathcal{L}}{\delta \left(
\partial _{+}\partial _{+}\phi ^{a}\right) }\ ,  \label{pi}
\end{eqnarray}
where the fields $\phi $ and $\partial _{+}\phi $ are treated as independent
canonical fields.

It is straightforward to show that the EL equations can be written by
\end{subequations}
\begin{equation*}
W_{ab}\left( \partial _{+}\right) ^{4}\phi ^{b}=F_{a}\left( \phi ,\partial
\phi ,\partial ^{2}\phi ,\partial ^{3}\phi \right)
\end{equation*}%
where the generalized Hessian matrix is%
\begin{equation}
W_{ab}\equiv \frac{\delta \pi _{a}}{\delta \left( \partial _{+}\partial
_{+}\phi ^{b}\right) }=\int d^{3}x\frac{\delta \mathcal{L}}{\delta \left(
\partial _{+}\partial _{+}\phi ^{a}\right) \delta \left( \partial
_{+}\partial _{+}\phi ^{b}\right) }\ .  \label{hessian}
\end{equation}%
It is the regularity or the singularity of this matrix that determines the
regularity or the singularity of the system.

In this analysis we have ignored the boundary conditions of the fields,
which is a quite misleading attitude, since the null-plane dynamics requires
a different analysis of initial-boundary conditions than the instant-form
dynamics. The discussion about the initial-boundary value problem in this
case will be made properly during the canonical procedure, so at this point
we just make sure that the conditions of the fields are equivalent of those
in instant-form, in other words, that the fields and all required
derivatives go to zero at the boundary of the 3-surface.

\section{The Hamiltonian analysis}

From the Lagrangian density (\ref{I1}) and the definitions (\ref{momenta})
follows the canonical momenta for the Podolsky's field
\begin{subequations}
\label{momenta1}
\begin{eqnarray}
p^{\mu } &=&F^{\mu +}-a^{2}\left( \eta ^{\mu -}\partial _{-}\partial
_{\lambda }F^{+\lambda }+\eta ^{\mu i}\partial _{i}\partial _{\lambda
}F^{+\lambda }-2\partial _{-}\partial _{\lambda }F^{\mu \lambda }\right) \ ,
\label{pmu} \\
\pi ^{\mu } &=&a^{2}\eta ^{\mu +}\partial _{\lambda }F^{+\lambda }\ .
\label{pimu}
\end{eqnarray}
The Hessian matrix of this system is
\end{subequations}
\begin{equation*}
W^{\mu \nu }=\frac{\delta \pi ^{\mu }}{\delta \left( \partial _{+}\partial
_{+}A_{\nu }\right) }=-a^{2}\eta ^{\mu +}\delta _{-}^{\nu }\eta ^{++}=0\ .
\end{equation*}%
As we saw in the earlier section the fields $A_{\mu }$ and $\partial
_{+}A_{\mu }$ should be treated as independent variables. Therefore we will
use the notation $\bar{A}_{\mu }\equiv \partial _{+}A_{\mu }$, being $A_{\mu
}$ and $\bar{A}_{\mu }$ independent fields. Then we are able to define the
primary constraints
\begin{subequations}
\label{11}
\begin{eqnarray}
\phi _{1} &=&\pi ^{+}\approx 0\ , \\
\phi _{2}^{i} &=&\pi ^{i}\approx 0\ , \\
\phi _{3} &=&p^{+}-\partial _{-}\pi ^{-}\approx 0\ , \\
\phi _{4}^{i} &=&p^{i}-\partial _{i}\pi ^{-}+F_{i-}+2a^{2}\partial _{-}\left[
\partial _{i}\bar{A}_{-}-2\partial _{-}\bar{A}_{i}+\partial _{i}\partial
_{-}A_{+}-\partial _{j}F_{ij}\right] \approx 0\ .
\end{eqnarray}

The canonical Hamiltonian density can be expressed by
\end{subequations}
\begin{eqnarray}
\mathcal{H}_{c} &=&p^{\mu }\bar{A}_{\mu }+\pi ^{-}\left( \partial _{-}\bar{A}%
_{+}-\partial ^{i}\bar{A}_{i}+\partial ^{i}\partial _{i}A_{+}\right) -\frac{1%
}{2}\left( \bar{A}_{-}-\partial _{-}A_{+}\right) ^{2}-\left( \bar{A}%
_{i}-\partial _{i}A_{+}\right) F_{-i}  \notag \\
&&+\frac{1}{4}F_{ij}F^{ij}+\frac{1}{2}a^{2}\left( \partial _{i}\bar{A}_{-}-2\partial
_{-}\bar{A}_{i}+\partial _{i}\partial _{-}A_{+}-\partial _{j}F_{ij}\right)
^{2}\ .  \label{Hc}
\end{eqnarray}%
With the canonical Hamiltonian $H_{c}=\int d^{3}x\mathcal{H}_{c}(x)$ and the
primary constraints (\ref{11}) we build the primary Hamiltonian%
\begin{equation}
H_{P}\equiv H_{c}+\int d^{3}xu^{a}(x)\phi _{a}(x)\ ,\hspace{1cm}\left\{
a\right\} =\left\{ 1,2,3,4\right\} \ .  \label{HP}
\end{equation}

To proceed with the calculus of the consistency conditions we use the
primary Hamiltonian as generator of the $x^{+}$ evolution and define the
fundamental equal $x^{+}$\ Poisson Brackets with the expressions
\begin{equation}
\left\{ A_{\mu }(x),p^{\nu }(y)\right\} _{x^{+}=y^{+}}=\left\{ \bar{A}_{\mu
}(x),\pi ^{\nu }(y)\right\} _{x^{+}=y^{+}}=\delta _{\mu }^{\nu }\delta
^{3}(x-y)\ ,  \label{PB1}
\end{equation}%
where $\delta ^{3}(x-y)\equiv \delta (x^{-}-y^{-})\delta ^{2}(\mathbf{x}-%
\mathbf{y})$. We verify that the condition $\dot{\phi}_{1}\approx 0$ gives
just the constraint $\phi _{3}\approx 0$, which is already satisfied. The
consistency for the remaining constraints gives equations for some Lagrange
multipliers. Notice that the conditions for $\phi _{2}^{i}$ and $\phi _{3}$,
\begin{eqnarray*}
\dot{\phi}_{2}^{i} &=&-\phi _{4}^{i}+4a^{2}\partial _{-}\partial
_{-}u_{i}^{4}\approx 0\ , \\
\dot{\phi}_{3} &=&\partial _{-}p^{-}+\partial _{i}p^{i}+4a^{2}\partial
_{i}\partial _{-}\partial _{-}u_{i}^{4}\approx 0\ ,
\end{eqnarray*}%
give equations for the same parameters $u_{i}^{4}$. These equations must be
consistent to each other. From the first we have $\partial _{-}\partial
_{-}u_{i}^{4}\approx 0$ and, applying this result on the second condition, a
secondary constraints appears:%
\begin{equation*}
\chi \equiv \partial _{-}p^{-}+\partial _{i}p^{i}\approx 0\ .
\end{equation*}%
For this secondary constraint, $\dot{\chi}=0$, and no more constraints can
be found.

The analysis leaves us with the following set:%
\begin{eqnarray*}
\chi &=&\partial _{-}p^{-}+\partial _{i}p^{i}\approx 0\ , \\
\phi _{1} &=&\pi ^{+}\approx 0\ , \\
\phi _{2}^{i} &=&\pi ^{i}\approx 0\ , \\
\phi _{3} &=&p^{+}-\partial _{-}\pi ^{-}\approx 0\ , \\
\phi _{4}^{i} &=&p^{i}-\partial _{i}\pi ^{-}+F_{i-}+2a^{2}\partial _{-}\left[
\partial _{i}\bar{A}_{-}-2\partial _{-}\bar{A}_{i}+\partial _{i}\partial
_{-}A_{+}-\partial _{j}F_{ij}\right] \approx 0\ .
\end{eqnarray*}%
It happens that $\chi $ and $\phi _{1}$ are first-class constraints, while $%
\phi _{2}^{i}$ , $\phi _{3}$ and $\phi _{4}^{i}$ are second-class ones.
However, constructing the matrix of the second-class constraints we found
that it is singular of rank four, which indicates that there must exist a
first-class constraint, associated with the zero mode of this matrix, and
its construction is made from the corresponding eigenvector which gives a
linear combination of second-class constraints. The combination happens to
be just $\Sigma _{2}\equiv \phi _{3}-\partial _{i}\phi _{2}^{i}$ and it is
independent of $\chi $ and $\phi _{1}$. Therefore, we have the renamed set
of first-class constraints
\begin{subequations}
\label{fcc}
\begin{eqnarray}
\Sigma _{1} &\equiv &\pi ^{+}\approx 0\ ,  \label{s1} \\
\Sigma _{2} &\equiv &p^{+}-\partial _{-}\pi ^{-}-\partial _{k}\pi
^{k}\approx 0\ ,  \label{s2} \\
\Sigma _{3} &\equiv &\partial _{-}p^{-}+\partial _{i}p^{i}\approx 0\ ,
\label{s3}
\end{eqnarray}%
and a set of irreducible second-class constraints
\end{subequations}
\begin{subequations}
\label{scc}
\begin{eqnarray}
\Phi _{1}^{i} &\equiv &\pi ^{i}\approx 0\ ,  \label{p1} \\
\Phi _{2}^{i} &\equiv &p^{i}-\partial _{i}\pi ^{-}+F_{i-}+2a^{2}\partial
_{-} \left[ \partial _{i}\bar{A}_{-}-2\partial _{-}\bar{A}_{i}+\partial
_{i}\partial _{-}A_{+}-\partial _{j}F_{ij}\right] \approx 0\ .  \label{p2}
\end{eqnarray}%
The second-class constraints do not appear in the instant-form dynamics for
this theory: they are a common effect of the null-plane dynamics.

Here we are in position to write the total Hamiltonian
\end{subequations}
\begin{equation}
H_{T}\equiv H_{c}+\int d^{3}xu^{a}(x)\Sigma _{a}(x)+\int d^{3}x\lambda
_{i}^{I}(x)\Phi _{I}^{i}(x)\ ,  \label{HT}
\end{equation}%
with which we are able to calculate the canonical equations of the system
for the variables $A_{\mu }$, $\bar{A}_{\mu }$, $p^{\mu }$ and $\pi ^{\mu }$.

For $A_{\mu }$ we have the equations%
\begin{equation}
\partial _{+}A_{\mu }=\bar{A}_{\mu }+\delta _{\mu }^{+}u^{2}-\delta _{\mu
}^{-}\partial _{-}u^{3}-\delta _{\mu }^{i}\left[ \partial _{i}u^{3}-\lambda
_{i}^{2}\right] \ ,  \label{A}
\end{equation}%
which just means that the canonical variable $\bar{A}_{\mu }$ is defined as $%
\partial _{+}A_{\mu }$ plus a linear combination of the still arbitrary
Lagrange multipliers. The equations for $\bar{A}_{\mu }$ give%
\begin{equation}
\partial _{+}\bar{A}_{\mu }\approx \delta _{\mu }^{+}u^{1}+\delta _{\mu }^{-}%
\left[ \partial _{-}\bar{A}_{+}+\partial _{i}\bar{A}_{i}-\partial
_{i}\partial _{i}A_{+}+\partial _{-}u^{2}+\partial _{i}\lambda _{i}^{2}%
\right] +\delta _{\mu }^{i}\left[ \partial _{i}u^{2}+\lambda _{i}^{1}\right]
\ .  \label{B}
\end{equation}%
The equation for $\bar{A}_{+}$ is just $\partial _{+}\bar{A}_{+}\approx
u^{1} $, which is expected since $\bar{A}_{+}$ is a degenerate variable. The
expression for $\bar{A}_{-}$ can be written, using (\ref{A}), as%
\begin{equation}
\partial _{\mu }F^{-\mu }\approx -\left[ \partial ^{i}\partial _{i}+\partial
^{+}\partial _{+}\right] u^{3}\ .  \label{B-}
\end{equation}

The Hamiltonian equations for the momenta $p^{\mu }$ are given, with (\ref{A}%
) and $\pi ^{-}=a^{2}\partial _{\lambda }F^{+\lambda }$, by%
\begin{eqnarray*}
\partial _{+}p^{+} &\approx &\partial _{\lambda }F^{\lambda
+}-a^{2}\partial _{i}\partial _{-}\partial _{\lambda }F^{\lambda
i}-a^{2}\partial _{i}\partial _{i}\partial _{\lambda }F^{\lambda +}+\left(
1+a^{2}\partial _{i}\partial _{i}\right) \partial _{-}\partial _{-}u^{3}\ ,
\\
\partial _{+}p^{-} &\approx &\partial _{i}F^{i-}+\partial _{i}\partial
_{i}u^{3}\ , \\
\partial _{+}p^{i} &\approx &\partial _{-}F^{-i}+\partial
_{j}F^{ji}-a^{2}\partial _{\mu }\partial ^{\mu }\partial
_{j}F^{ij}-\partial _{-}\partial _{i}u^{3}\ .
\end{eqnarray*}%
The equations for $\pi ^{\mu }$ are, using the fact that $\pi ^{+}$ and $\pi
^{i}$ are weakly zero,%
\begin{eqnarray*}
p^{+} &\approx &a^{2}\partial _{-}\partial _{\lambda }F^{+\lambda }\ , \\
p^{-} &\approx &F^{-+}+a^{2}\partial _{-}\partial _{\lambda }F^{-\lambda
}+\partial _{-}u^{3}-a^{2}\partial _{-}\partial _{i}\partial _{i}u^{3}\ , \\
p^{i} &\approx &F^{i+}-a^{2}\left( \partial ^{i}\partial _{\lambda
}F^{+\lambda }-2\partial _{-}\partial _{\lambda }F^{i\lambda }\right)
+2a^{2}\partial _{-}\partial _{-}\partial _{i}u^{3}\ .
\end{eqnarray*}%
The last equations reproduce the definition of the canonical momenta $p$
with some combination of the Lagrange multipliers. If we use these equation
on the earlier equations for $\partial _{+}p^{\mu }$, and also using (\ref%
{B-}), we have
\begin{subequations}
\begin{eqnarray}
\left( 1+a^{2}\square \right) \partial _{\lambda }F^{\lambda +}+\left(
1+a^{2}\partial _{i}\partial _{i}\right) \partial _{-}\partial _{-}u^{3}
&\approx &0\ ,  \label{ce1} \\
\left( 1+a^{2}\square \right) \partial _{\lambda }F^{\lambda
-}+a^{2}\partial _{+}\partial _{-}\partial _{i}\partial _{i}u^{3} &\approx
&0\ ,  \label{ce2} \\
\left( 1+a^{2}\square \right) \partial _{\lambda }F^{\lambda i}-\left(
1+2a^{2}\partial _{+}\partial _{-}\right) \partial _{-}\partial _{i}u^{3}
&\approx &0\ .  \label{ce3}
\end{eqnarray}%
These equations are compatible with the Lagrangian field equations (\ref%
{ELPodolsky}) only if suitable gauge conditions are chosen in order to
eliminate the Lagrange multiplier $u^{3}$.

\section{Gauge fixing and Dirac Brackets}

At this stage we have a set of first-class constraints, the relations (\ref%
{fcc}), that must be considered as generators of gauge transformations. The
problem of choosing proper gauge conditions has to be solved to fully
eliminate the redundant variables of the theory at the classical level and,
therefore, to proceed with a consistent quantization of the Podolsky's field.

As it has already stated in the introduction section, the first attempt to
find gauge conditions in the instant-form of the theory was made by using
the Lorenz gauge $\partial _{\mu }A^{\mu }=0$. However, as showed in \cite{8}%
, the Lorenz condition is not a good gauge choice for the Podolsky's field,
since it does not fulfill the necessary requirements for a consistent gauge:
it does not fix the gauge, it is not preserved by the equations of motion
and it is not attainable. Moreover, it is also clear that the solutions of
the field equations (\ref{ELPodolsky}) cannot consist only by transverse
fields.

The analysis of the correct gauge fixing on the null-plane can be made by
closely inspect the EL equations of the system. If we look for the $\mu =+$
equation, it produces the explicit solution for $A_{+}$%
\end{subequations}
\begin{equation}
A_{+}=-\frac{1}{\left( 1+a^{2}\square \right) \nabla ^{2}}\partial
_{+}\left( 1+a^{2}\square \right) \left( \partial ^{-}A_{-}+\partial
^{i}A_{i}\right) \ ,  \label{Ap}
\end{equation}%
where $\nabla ^{2}\equiv \partial _{i}\partial _{i}$. The remaining
equations of motion can be written, eliminating the $A_{+}$ variable, by%
\begin{equation*}
\left( 1+a^{2}\square \right) \square \mathbf{A}_{-}=0\ ,\ \ \ \ \ \ \ \ \
\ \ \left( 1+a^{2}\square \right) \square \mathbf{A}_{i}=0\ ,
\end{equation*}%
with%
\begin{eqnarray*}
\mathbf{A}_{-} &\equiv &A_{-}+\partial _{-}\frac{1}{\left( 1+a^{2}\square
\right) \nabla ^{2}}\left( 1+a^{2}\square \right) \left( \partial
^{-}A_{-}+\partial ^{i}A_{i}\right) \ , \\
\mathbf{A}_{i} &\equiv &A_{i}+\partial _{i}\frac{1}{\left( 1+a^{2}\square
\right) \nabla ^{2}}\left( 1+a^{2}\square \right) \left( \partial
^{-}A_{-}+\partial ^{j}A_{j}\right) \ .
\end{eqnarray*}%
Therefore, we can achieve the variables $\mathbf{A}$ through a gauge
transformation such that the gauge function is%
\begin{equation*}
\Delta =\frac{1}{\left( 1+a^{2}\square \right) \nabla ^{2}}\left(
1+a^{2}\square \right) \left( \partial ^{-}A_{-}+\partial ^{i}A_{i}\right)
\ .
\end{equation*}%
In addition, these fields satisfy the condition%
\begin{equation}
\left( 1+a^{2}\square \right) \left( \partial ^{-}\mathbf{A}_{-}+\partial
^{i}\mathbf{A}_{i}\right) =0\ ,  \label{coul}
\end{equation}%
which is the generalized Coulomb condition on the null-plane.

For that reason, the most natural gauge choice that is compatible with the
field equations is given by%
\begin{equation}
\left( 1+a^{2}\square \right) \left( \bar{A}_{-}+\partial ^{i}A_{i}\right)
\approx 0\ .  \label{gauge}
\end{equation}%
Back to (\ref{Ap}) we see that the time preservation of this relation is
guaranteed if we set $A_{+}\approx 0$. Whereas, consistency requires $\bar{A}%
_{+}\approx 0$ as well.

In this gauge the field equations are written by%
\begin{equation*}
\left( 1+a^{2}\square \right) \square A_{B}=0\ ,
\end{equation*}%
which is a generalized wave equation on the null-plane for the variables $%
A_{B}\equiv \left( A_{-},A_{i}\right) $.

Back to the Hamiltonian framework, this analysis leads to the gauge
conditions
\begin{subequations}
\label{g1}
\begin{eqnarray}
\Omega _{1} &\equiv &\bar{A}_{+}\approx 0\ ,  \label{o1} \\
\Omega _{2} &\equiv &A_{+}\approx 0\ ,  \label{o2} \\
\Omega _{3} &\equiv &\left( 1+a^{2}\square \right) \left( \bar{A}%
_{-}+\partial ^{i}A_{i}\right) \approx 0\ ,  \label{o3}
\end{eqnarray}%
which is the generalized radiation gauge on the null-plane. The next step is
to calculate Dirac Brackets for the set of ten constraints of the theory,
but due to the present of the second-class constraints (\ref{scc}) it is
more convenient to evaluate the reduced dynamics for these constraints
first. Taking the matrix of the Poisson Brackets of the second-class
constraints we have
\end{subequations}
\begin{equation}
M^{ij}\equiv 2\eta ^{ij}\partial _{-}^{x}\left(
\begin{array}{cc}
0 & -2a^{2}\partial _{-}^{x} \\
2a^{2}\partial _{-}^{x} & 1-2a^{2}\nabla _{x}^{2}%
\end{array}%
\right) \delta ^{3}(x-y)\ .  \label{M}
\end{equation}%
The explicit evaluation of the inverse involves the knowledge of the inverse
of the operators $\left( \partial _{-}^{x}\right) ^{-1}$, $\left( \partial
_{-}^{x}\right) ^{-2}$, and $\left( \partial _{-}^{x}\right) ^{-3}$, which
are Green's functions of the operators $\partial _{-}^{x}$, $\left( \partial
_{-}^{x}\right) ^{2}$, and $\left( \partial _{-}^{x}\right) ^{3}$. To
achieve a unique solution it is necessary and sufficient to impose $\partial
_{-}^{x}A_{\mu }=0$, $\partial _{-}^{x}\partial _{-}^{x}A_{\mu }=0$, and $%
\partial _{-}^{x}\partial _{-}^{x}\partial _{-}^{x}A_{\mu }=0$ on $%
x^{+}\rightarrow -\infty $ as the appropriate initial conditions of the
theory. This choice is also consistent with the definition of momenta (\ref%
{momenta}), since their definition are also dependent on initial-boundary
conditions. Therefore, we write the unique inverse%
\begin{equation}
N_{ij}\left( x,y\right) \equiv \frac{1}{2}\eta _{ij}\left(
\begin{array}{cc}
\alpha \left( x,y\right) & \beta \left( x,y\right) \\
\gamma \left( x,y\right) & 0%
\end{array}%
\right)  \label{N}
\end{equation}%
with the coefficients
\begin{subequations}
\label{alphabeta}
\begin{eqnarray}
\alpha \left( x,y\right) &=&\frac{1}{4a^{4}}\left( x^{-}-y^{-}\right)
^{2}\epsilon \left( x^{-}-y^{-}\right) \left( 1-2a^{2}\nabla _{x}^{2}\right)
\delta ^{2}(\mathbf{x}-\mathbf{y})\ ,  \label{alpha} \\
\beta \left( x,y\right) &=&-\gamma (x,y)=\frac{1}{a^{2}}\ \left\vert
x^{-}-y^{-}\right\vert \delta ^{2}(\mathbf{x}-\mathbf{y})\ .  \label{beta}
\end{eqnarray}

With this inverse we are able to define the first Dirac Brackets for two
observables $A(x)$ and $B(y)$,
\end{subequations}
\begin{equation}
\left\{ A(x),B(y)\right\} ^{\ast }=\left\{ A(x),B(y)\right\} -\iint
d^{3}zd^{3}w\left\{ A(x),\Phi _{I}^{i}(z)\right\} N_{ij}^{IJ}(z,w)\left\{
\Phi _{J}^{j}(w),B(y)\right\} \ ,  \label{1DB}
\end{equation}%
where $\left\{ I,J\right\} =\left\{ 1,2\right\} $. This definition implies
elimination of the second-class constrains and the definition of an extended
Hamiltonian where $\Phi _{I}$ are strongly zero. Thus, we are left with the
first-class constraints $\Sigma $ and the gauge conditions $\Omega $. To
proceed with the evaluation of the complete Dirac Brackets we should
calculate the matrix of the first Dirac Brackets of these constraints. It
is given by%
\begin{equation*}
C(x,y)\equiv \left\{ \chi _{A}(x),\chi _{B}(y)\right\} ^{\ast }=\left(
\begin{array}{cc}
0 & \mathcal{O}(x,y) \\
-\mathcal{O}^{T}(x,y) & 0%
\end{array}%
\right)
\end{equation*}%
with $\chi _{A}\equiv \left( \Sigma _{a},\Omega _{a}\right) $. If we write $%
\mathcal{D}_{x}\equiv \left( 1-a^{2}\nabla _{x}^{2}\right) $ the matrix $%
\mathcal{O}$ follows:
\begin{equation*}
\mathcal{O}(x,y)=\left(
\begin{array}{ccc}
-1 & 0 & 0 \\
0 & -1 & \mathcal{D}_{x}\partial _{-}^{x} \\
0 & 0 & -\mathcal{D}_{x}\nabla _{x}^{2}%
\end{array}%
\right) \delta ^{3}(x-y)\ .
\end{equation*}%
The inverse is given by%
\begin{equation}
C^{-1}(x,y)=\left(
\begin{array}{cc}
0 & -\left( \mathcal{O}^{-1}\right) ^{T}(x,y) \\
\mathcal{O}^{-1}(x,y) & 0%
\end{array}%
\right) \ ,  \label{Cinv}
\end{equation}%
in which%
\begin{equation}
\mathcal{O}^{-1}(x,y)=\left(
\begin{array}{ccc}
-\delta ^{3}(x-y) & 0 & 0 \\
0 & -\delta ^{3}(x-y) & \gamma \left( x,y\right) \\
0 & 0 & \rho \left( x,y\right)%
\end{array}%
\right) \ .  \label{Oinv}
\end{equation}

Under the considered boundary conditions the coefficients are given by
\begin{subequations}
\begin{eqnarray}
\gamma \left( x,y\right) &=&-\partial _{-}^{x}\left( \nabla _{x}^{2}\right)
^{-1}\delta ^{3}(x-y)\ ,  \label{i1} \\
\rho \left( x,y\right) &=&-\mathcal{D}_{x}^{-1}\left( \nabla _{x}^{2}\right)
^{-1}\delta ^{3}(x-y)\ ,  \label{i2}
\end{eqnarray}%
where
\end{subequations}
\begin{equation*}
\left( \nabla _{x}^{2}\right) ^{-1}\delta ^{3}(x-y)=\frac{1}{4\pi }\ln
\left( \mathbf{x}-\mathbf{y}\right) ^{2}.
\end{equation*}%
Then, we are able to define the complete Dirac Brackets of the generalized
radiation gauge:%
\begin{eqnarray}
\left\{ A(x),B(y)\right\} ^{\ast \ast } &\equiv &\left\{ A(x),B(y)\right\}
^{\ast }+\int d^{3}zd^{3}w\left\{ A(x),\Sigma _{a}(z)\right\} ^{\ast }\left[
\left( \mathcal{O}^{-1}\right) ^{T}\right] ^{ab}(z,w)\left\{ \Omega
_{b}(w),B(y)\right\} ^{\ast }  \notag \\
&&-\int d^{3}zd^{3}w\left\{ A(x),\Omega _{a}(z)\right\} ^{\ast }\left[
\mathcal{O}^{-1}\right] ^{ab}(z,w)\left\{ \Sigma _{b}(w),B(y)\right\} ^{\ast
}\ .  \label{Dirbra}
\end{eqnarray}%
The complete set of fundamental DB follows:%
\begin{eqnarray*}
\left\{ A_{\mu }(x),\bar{A}_{\nu }(y)\right\} ^{\ast \ast } &=&-\frac{1}{%
8a^{2}}\delta _{\mu }^{i}\delta _{\nu }^{j}\eta _{ij}\left\vert
x^{-}-y^{-}\right\vert \delta ^{2}(\mathbf{x}-\mathbf{y}), \\
\left\{ A_{\mu }(x),p^{\nu }(y)\right\} ^{\ast \ast } &=&\left[ \delta _{\mu
}^{\nu }-\delta _{\mu }^{+}\delta _{+}^{\nu }+\left( \delta _{\mu
}^{-}\partial _{-}+\delta _{\mu }^{i}\partial _{i}\right) \left( \delta
_{+}^{\nu }\partial _{-}+\delta _{j}^{\nu }\partial ^{j}\right) \frac{1}{%
\nabla ^{2}}\right] \delta ^{3}(x-y), \\
\left\{ A_{\mu }(x),\pi ^{\nu }(y)\right\} ^{\ast \ast } &=&\delta _{-}^{\nu
}\left[ \delta _{\mu }^{-}\partial _{-}+\delta _{\mu }^{i}\partial _{i}%
\right] \frac{1}{\nabla ^{2}}\delta ^{3}(x-y), \\
\left\{ \bar{A}_{\mu }(x),\bar{A}_{\nu }(y)\right\} ^{\ast \ast } &=&\frac{1%
}{64a^{4}}\delta _{\mu }^{i}\delta _{\nu }^{j}\eta _{ij}\left(
x^{-}-y^{-}\right) ^{2}\epsilon \left( x^{-}-y^{-}\right) \left[
1-2a^{2}\nabla ^{2}\right] \delta ^{2}(\mathbf{x}-\mathbf{y}) \\
&&-\frac{1}{8a^{2}}\eta _{ij}\left\vert x^{-}-y^{-}\right\vert \left[
\delta _{\mu }^{i}\delta _{\nu }^{-}+\delta _{\mu }^{-}\delta _{\nu }^{i}%
\right] \partial _{j}\delta ^{2}(\mathbf{x}-\mathbf{y}), \\
\left\{ \bar{A}_{\mu }(x),p^{\nu }(y)\right\} ^{\ast \ast } &=&\frac{1}{%
8a^{2}}\delta _{\mu }^{i}\delta _{-}^{\nu }\left\vert
x^{-}-y^{-}\right\vert \partial _{i}\delta ^{2}(\mathbf{x}-\mathbf{y}) \\
&&-\frac{1}{4}\delta _{\mu }^{i}\epsilon \left( x^{-}-y^{-}\right) \left[
\delta _{k}^{\nu }\partial _{k}\partial _{i}+\frac{1}{2a^{2}}\delta
_{i}^{\nu }\left( 1-2a^{2}\nabla ^{2}\right) \right] \delta ^{2}(\mathbf{x}-%
\mathbf{y}) \\
&&-\delta _{+}^{\nu }\left[ \left[ \delta _{\mu }^{-}\partial _{-}+\delta
_{\mu }^{i}\partial _{i}\right] -\frac{1}{2}\delta _{\mu }^{i}\partial _{i}%
\right] \delta ^{3}(x-y), \\
\left\{ \bar{A}_{\mu }(x),\pi ^{\nu }(y)\right\} ^{\ast \ast } &=&\left[
\delta _{\mu }^{\nu }-\delta _{\mu }^{+}\delta _{+}^{\nu }+\eta _{\mu
j}\delta _{j}^{\nu }\right] \delta ^{3}(x-y)-\frac{1}{4}\eta _{\mu j}\delta
_{-}^{\nu }\epsilon \left( x^{-}-y^{-}\right) \partial _{j}\delta ^{2}(%
\mathbf{x}-\mathbf{y}).
\end{eqnarray*}

With these brackets we can deduce the fundamental ones that will lead,
through the correspondence principle, to a consistent quantization of the
field. The physical degrees of freedom can be found with the analysis of the
constraints as strong relations. Of course, the fields $A_{+}$, $\bar{A}_{+}$%
, $\pi ^{+}$ and $\pi ^{i}$ are not independent, since they are strongly
zero in the formalism. Thus, $p^{i}$, $p^{+}$ and $p^{-}$ can be written in
function of $\pi ^{-}$ and other variables. The gauge condition (\ref{o3})
also eliminates $\bar{A}_{-}$. Therefore, the only independent variables are
actually given by $A_{-},A_{i},\bar{A}_{i},p^{i}$ and $\pi ^{-}$. They are
eight independent fields, less than the dynamics in instant-form \cite{8},
which can be seen as a good feature, but the structure of the phase space
comes out to be quite more complicate.

Considering, for example, the brackets%
\begin{equation}
\left\{ A_{-}(x),p^{i}(y)\right\} ^{\ast \ast }=\frac{\partial _{-}\partial
^{i}}{\nabla ^{2}}\delta ^{3}(x-y),
\end{equation}%
we can see that the longitudinal field acquires a non-local character. This
is expected for every system analyzed on the null-plane, since this
component lies on the light-cone and no criterium of causality can be
employed for this field. The non-locality is due to the second-class
constraints, which does not appear in instant-form dynamics of the system.

For the transverse fields the brackets
\begin{equation}
\left\{ A_{i}(x),p^{j}(y)\right\} ^{\ast \ast }=\left[ \delta _{i}^{j}-\frac{%
\partial _{i}\partial _{j}}{\nabla ^{2}}\right] \delta ^{3}(x-y)
\end{equation}%
indicates that a Coulomb-type interaction is present, this case in two
dimensions, what justifies to call the gauge condition (\ref{gauge}) the
generalized Coulomb condition. This is also expected, since these brackets
depends exclusively on the first-class constraints plus gauge conditions,
just like in the instant-form.

\section{Final remarks}

We have analysed the canonical structure of Podolsky's electrodynamics on
the null-plane. The theory has high-order derivatives in the Lagrangian
function, so we followed the procedure outlined in \cite{8} for the
definition of the Hamiltonian density (\ref{energy}) and of the canonical
momenta associated to the fields $A_{\mu }$ and $\bar{A}_{\mu }$ (\ref%
{momenta}), which result from the definition of the conserved
energy-momentum tensor.

We have observed in the study of the initial-boundary problem of Podolsky's
equation that, because it is a second-order equation, the uniqueness of the
solution is obtained when the field $A_{\mu }$ is specified on the
null-plane $x^{+}=cte$ and three boundary conditions are imposed on $%
x^{-}=cte$. These conditions where chosen to be $\partial _{-}A_{\mu }=0$, $%
\left( \partial _{-}\right) ^{2}A_{\mu }=0$, and $\left( \partial
_{-}\right) ^{3}A_{\mu }=0$ on $x^{-}\rightarrow -\infty $.

In the canonical analysis of the Podolsky's theory we found a set of three
first-class constraints (\ref{fcc}), and a set of four
second-class ones (\ref{scc}). The first-class constraints are responsible
for the $U\left( 1\right) $ invariance of the Action, which is expected
since the gauge character of the field should not be destroyed by the choice
of parametrization. The form of this set is analogous to the set found in
instant-form \cite{8}, which is also expected.

The new feature on the null-plane is the second-class constraints, which are
not present in the conventional instant-form dynamics \cite{8}. The
appearance of second-class constraints is a common effect of the null-plane
dynamics \cite{ste,Cas,9,10}, and they are responsible to the fact that the
analysis on the null-plane requires a lesser number of degrees of freedom.
Because of the second-class constraints the longitudinal components of the
fields turned out to be non-local.

To evaluate the physical degrees of freedom it was necessary to choose
proper gauge conditions for the theory, which was a subject that needed
closer inspection. Gauge conditions must obey a set of requirements to be
consistent with the formalism: they must fix completely the gauge, they must
be consistent with the field equations, they must not affect Lorentz
covariance and, last but not least, they must be attainable. Therefore, we
followed the procedure outlined in \cite{8} and found that the generalized
radiation gauge (\ref{g1}) on the null-plane fulfill all these requirements.
Of course, this gauge choice is not the only consistent possible choice.
There is, for example, the so called null-plane gauge, which will be studied
in a future work concerning the Podolsky's field coupled with scalar and
spinor fields.

Since the first and second-class constraints, together with the gauge
conditions were known, we calculated the Dirac Brackets that had clarified
the physical fields of the system. However, these brackets are not unique
unless we specify all the information about the initial-boundary value
problem of the theory. By imposing the value of the field on the null-plane $%
x^{+}=cte$, and the considered boundary conditions on $x^{-}=cte$ , we have
fixed the hidden subset of the first-class constraints \cite{ste,BCT} and
got a unique inverse for the second-class constraints matrix when the
ambiguity on the operators $\left( \partial _{-}^{x}\right) ^{-1}$, $\left(
\partial _{-}^{x}\right) ^{-2}$, and $\left( \partial _{-}^{x}\right) ^{-3}$
was eliminated.

Finally, an analysis of the physical fields results in the true degrees of
freedom, which are given by $A_{-},A_{i},\bar{A}_{i},p^{i}$ and $\pi ^{-}$.
The complete Dirac Brackets of these fields implicated the non-locality of
the longitudinal component $A_{-}$ and a Coulomb-type interaction in the
electrostatic case, in two dimensions.

\section*{ACKNOWLEDGEMENTS}

The authors would like to thank Professor C.A.P. Galv\~{a}o for reading the
manuscript and to contribute with the improvement of our work. M.C. Bertin
was supported by Capes. B.M. Pimentel was partially supported by CNPq.
G.E.R. Zambrano was supported by CNPq.

\end{document}